\documentclass[11pt,a4paper]{article}

\usepackage{amssymb}

\usepackage[dvips]{graphicx}

\begin{document}

\title{\bf Relation between two-point Green functions of ${\cal N}=1$ SQED with $N_f$
flavors, regularized by higher derivatives, in the three-loop
approximation}

\author{
A.E.Kazantsev, K.V.Stepanyantz\\
{\small{\em Moscow State University}}, {\small{\em  Physical
Faculty, Department  of Theoretical Physics}}\\
{\small{\em 119991, Moscow, Russia}}}

\maketitle

\begin{abstract}
We verify the identity which relates the two-point Green functions
of ${\cal N}=1$ SQED with $N_f$ flavors, regularized by higher
derivatives, by explicit calculations in the three-loop
approximation. This identity explains why in the limit of the
vanishing external momentum the two-point Green function of the
gauge superfield is given by integrals of double total derivatives
in the momentum space. It also allows to derive the NSVZ
$\beta$-function exactly in all loops if the renormalization group
functions are defined in terms of the bare coupling constant. In
order to verify the considered identity we use it for constructing
integrals giving the three-loop $\beta$-function starting from the
two-point Green functions of the matter superfields in the
two-loop approximation. Then we demonstrate that the results for
these integrals coincide with the sums of the corresponding
three-loop supergraphs.
\end{abstract}

\unitlength=1cm



\section{Introduction}
\hspace{\parindent}

The NSVZ $\beta$-function is a non-trivial relation between the
$\beta$-function of ${\cal N}=1$ supersymmetric gauge theories and
the anomalous dimensions of the matter superfields. For the pure
${\cal N}=1$ supersymmetric Yang-Mills theory it gives the exact
in all orders expression for the $\beta$-function, which appears
to be a geometric progression. The NSVZ $\beta$-function was
obtained using various general arguments, namely, the structure of
instanton contributions
\cite{Novikov:1983uc,Novikov:1985rd,Shifman:1999mv}, the anomaly
supermultiplet
\cite{Jones:1983ip,Shifman:1986zi,ArkaniHamed:1997mj},
nonrenormalization of the topological term \cite{Kraus:2002nu}.
Analogs of the NSVZ $\beta$-function were also found in theories
with softly broken supersymmetry for the renormalization of the
gaugino mass \cite{Hisano:1997ua,Jack:1997pa,Avdeev:1997vx} and
for ${\cal N}=(0,2)$ deformed $(2,2)$ two-dimensional sigma models
\cite{Chen:2014efa}.

In this paper we consider the ${\cal N}=1$ supersymmetric
electrodynamics (SQED) with $N_f$ flavors, for which the NSVZ
$\beta$-function is given by

\begin{equation}\label{NSVZ}
\beta(\alpha) = \frac{\alpha^2
N_f}{\pi}\Big(1-\gamma(\alpha)\Big),
\end{equation}

\noindent where $\gamma(\alpha)$ is the anomalous dimension of the
chiral matter superfields, and $\alpha$ denotes the argument of
functions in this relation. In the Abelian case the NSVZ
$\beta$-function was first obtained in Refs.
\cite{Vainshtein:1986ja,Shifman:1985fi}. For Abelian theories the
derivation of the NSVZ relation by explicit summation of Feynman
diagrams was made in \cite{Stepanyantz:2011jy,Stepanyantz:2014ima}
in all orders for the RG functions defined in terms of the bare
coupling constant in the case of using the higher derivative
regularization.

If the RG functions are defined in terms of the renormalized
coupling constant, comparing the NSVZ $\beta$-function with
explicit calculations of Feynman (super)graphs one should take
into account the ambiguity of choosing the subtraction scheme
\cite{Vladimirov:1975mx}. In theories with a single coupling
constant the $\beta$-function is scheme-dependent starting from
the three-loop approximation. The anomalous dimension is
scheme-dependent starting from the two-loop approximation.
Therefore, terms proportional to $\alpha^4$ in both sides of the
NSVZ relation are scheme dependent. The subtraction scheme
producing the NSVZ relation (so called\, "NSVZ scheme") in all
orders in the Abelian case was constructed in Ref.
\cite{Kataev:2013eta} by the help of the higher derivative
regularization.

Nevertheless, the most popular procedure for making calculations
in supersymmetric theories is the dimensional reduction
\cite{Siegel:1979wq} complemented by the $\overline{\mbox{DR}}$
subtraction scheme. Using this method the NSVZ relation was
verified up to the four-loop approximation
\cite{Avdeev:1981ew,Jack:1996vg,Jack:1996cn,Harlander:2006xq,Jack:2007ni}
(see Ref. \cite{Mihaila:2013wma} for a recent review). The
scheme-independent terms proportional to $\alpha^2$ and $\alpha^3$
appeared to be the same. However, starting from the three-loop
approximation the calculations made in the
$\overline{\mbox{DR}}$-scheme give results which do not satisfy
the NSVZ relation. However, it is possible to find a finite
renormalization after that the NSVZ relation is restored
\cite{Jack:1996vg,Jack:1996cn,Jack:1998uj}. The possibility of
making this redefinition is nontrivial \cite{Jack:1996vg}. The
reason is that in higher loops some terms in the NSVZ relation are
scheme independent. In particular, for ${\cal N}=1$ SQED with
$N_f$ flavors the terms proportional to the first degree of $N_f$
are scheme-independent and should satisfy NSVZ relation in all
orders \cite{Kataev:2013csa}. In the non-Abelian case this is
valid for terms proportional to $\mbox{tr}\, \left(C(R)^L\right)$
in $L$ loops \cite{Kataev:2014gxa}. Thus, even in higher loops the
NSVZ relation imposes nontrivial constrains on the divergences.
Nevertheless, (at present) the calculations made with the
dimensional reduction do not allow to construct in all orders a
scheme in which the NSVZ relation takes place. Moreover, the
dimensional reduction (unlike the dimensional regularization
\cite{'tHooft:1972fi,Bollini:1972ui,Ashmore:1972uj,Cicuta:1972jf},
which breaks supersymmetry \cite{Delbourgo:1974az}) is not
mathematically consistent \cite{Siegel:1980qs}. The price for
removal of inconsistencies is the loss of explicit supersymmetry
\cite{Avdeev:1981vf}. This implies that supersymmetry can be
broken by quantum corrections in higher loops
\cite{Avdeev:1982np,Avdeev:1982xy}. The calculations made in
\cite{Avdeev:1982np} and subsequently corrected in
\cite{Velizhanin:2008rw} demonstrate this breaking in the
three-loop approximation for ${\cal N}=2$ supersymmetric
Yang--Mills theory (without hypermultiplets). Most other
regularizations (see, e.g., \cite{Shifman:1985tj,Mas:2002xh}) were
usually used only for calculations in the one- and two-loop
approximations, where the problems related to the scheme
dependence are not essential.

It appears that a very convenient regularization for calculations
of quantum corrections in supersymmetric theories is the higher
covariant derivative regularization proposed by A.A.Slavnov
\cite{Slavnov:1971aw,Slavnov:1972sq}. Unlike the dimensional
reduction, the regularization by higher covariant derivatives is
consistent. For ${\cal N}=1$ supersymmetric gauge theories it can
be formulated in a manifestly supersymmetric way
\cite{Krivoshchekov:1978xg,West:1985jx}. Thus, with this
regularization supersymmetry is not broken at any stage of quantum
correction calculation. The higher covariant derivative
regularization can be also formulated for ${\cal N}=2$
supersymmetric gauge theories
\cite{Krivoshchekov:1985pq,Buchbinder:2014wra}. At least in the
Abelian case using of the higher covariant derivative
regularization allowed to solve the long standing problem of
constructing a renormalization prescription which gives the NSVZ
scheme in all orders \cite{Kataev:2013eta}, see also
\cite{Kataev:2014gxa}. It was noted \cite{Kataev:2013eta} that one
should distinguish the RG functions defined in terms of the bare
coupling constant and the RG functions defined in terms of the
renormalized coupling constant. If a supersymmetric gauge theory
is regularized by higher derivatives, the NSVZ relation appears to
be valid for the RG function defined in terms of the bare coupling
constant \cite{Stepanyantz:2011jy,Stepanyantz:2014ima} (see Ref.
\cite{Kataev:2014gxa} for a brief review). These RG functions are
scheme-independent for a fixed regularization (see, e.g.,
\cite{Kataev:2013eta}), that is do not depend on an arbitrariness
in a choice of the renormalization constants. The definition of
the RG function in terms of the bare coupling constant was used in
a lot of early papers devoted to the NSVZ $\beta$-function
\cite{Novikov:1983uc,Novikov:1985rd,ArkaniHamed:1997mj,Vainshtein:1986ja,Shifman:1985fi,Shifman:1985vu}.
However, standardly, the RG functions are defined in terms of the
renormalized coupling constant. Such RG functions are scheme
dependent and satisfy the NSVZ relation only in the NSVZ scheme.
In case of using the $\overline{\mbox{DR}}$-scheme the NSVZ
relation can be obtained after a specially tuned finite
renormalization in every order
\cite{Jack:1996vg,Jack:1996cn,Jack:1998uj}, which, however, is not
so far constructed in all orders. The higher covariant derivative
regularization allows to construct the NSVZ scheme in all orders
by imposing the simple boundary conditions
\cite{Kataev:2013eta}\footnote{Note that in the three-loop
approximation the NSVZ scheme obtained with the higher derivative
regularization differs from the NSVZ scheme constructed in
\cite{Jack:1996vg}, because the expressions for the anomalous
dimension are different. Certainly, these two NSVZ schemes can be
related by a finite renormalization, which includes finite
renormalization of the chiral matter superfields.}

\begin{equation}\label{NSVZ_Boundary_Conditions}
Z(\alpha, x_0)=1;\qquad Z_3(\alpha,x_0)=1,
\end{equation}

\noindent where $x_0$ is an arbitrary fixed value of
$\ln\Lambda/\mu$, on the renormalization constants. They ensure
that the RG functions defined in terms of the renormalized
coupling constant coincide with the ones defined in terms of the
bare coupling constant and satisfy the NSVZ relation in all
orders.

Certainly, the main statement for constructing the NSVZ scheme is
that the RG functions defined in terms of the bare coupling
constant satisfy the NSVZ relation in all orders if higher
derivatives are used for regularization. It is based on the
observation that in this case the integrals giving the
$\beta$-function are integrals of total derivatives
\cite{Soloshenko:2003nc} and even double total derivatives
\cite{Smilga:2004zr} in the momentum space. In the Abelian case
this fact was proved exactly in all orders by two different
methods \cite{Stepanyantz:2011jy,Stepanyantz:2014ima}, which give
the same result. For a general non-Abelian ${\cal N}=1$
supersymmetric gauge theory with chiral matter superfields
factorization of integrands into total derivatives was explicitly
demonstrated in the two-loop approximation
\cite{Pimenov:2009hv,Stepanyantz:2011zz}. Subsequently it was
verified that the corresponding integrals are also integrals of
double total derivatives
\cite{Stepanyantz:2011bz,Stepanyantz:2012zz,Stepanyantz:2012us}.
Such a structure of integrals allows to calculate one of the
momentum integrals analytically and obtain the exact NSVZ
$\beta$-function. In this paper we would like to verify the
results of Ref. \cite{Stepanyantz:2014ima}, in which the NSVZ
relation is obtained by the method based on the Schwinger--Dyson
equations for ${\cal N}=1$ SQED with $N_f$ flavors. (The
Schwinger--Dyson equations are widely used in quantum field
theory, see, e.g.,
\cite{Bogolyubov:1980nc,Bashir:2012fs,Broadhurst:1995dq}. A
possibility of using them for the derivation of the NSVZ relation
was first discussed in \cite{Stepanyantz:2004sg}.) The
Schwinger--Dyson equations allow to construct an equality relating
two-point Green functions of the gauge superfield and of the
matter superfields. This relation enables us to obtain integrals
giving the $\beta$-function defined in terms of the bare coupling
constant in every order if the integrals giving the two-point
Green functions for the chiral matter superfields are known in the
previous order. (The integrals for the $\beta$-function appear to
be integrals of double total derivatives.) In this paper we verify
this relation in the three-loop approximation. Namely, starting
from the two-loop integrals for the two-point Green functions of
the matter superfields we obtain the three-loop $\beta$-function
in the form of integrals of double total derivatives in the
momentum space. Then we compare the result with the known
expressions for the sums of the three-loop supergraphs.

This paper is organized as follows: In Sect.
\ref{Section_Regularization} we recall basic information
concerning ${\cal N}=1$ SQED with $N_f$ flavors regularized by
higher derivatives and the relation between the two-point Green
functions in this theory. In the three-loop approximation this
identity is verified in Sect. \ref{Section_Explicit_Three_Loop} by
explicit calculation of the supergraphs.

\section{${\cal N}=1$ SQED with $N_f$ flavors, regularized
by higher derivatives, and the relation between its two-point
Green functions} \hspace{\parindent}\label{Section_Regularization}

Following the paper \cite{Stepanyantz:2014ima}, here we consider
${\cal N}=1$ SQED with $N_f$ flavors, which (in the massless
limit) is described in terms of superfields by the following
action \cite{West:1990tg,Buchbinder:1998qv}:

\begin{equation}\label{N=1SQED}
S = \frac{1}{4e_0^2}\mbox{Re}\int d^4x\,d^2\theta\,W^a W_a +
\frac{1}{4} \sum\limits_{\alpha=1}^{N_f} \int
d^4x\,d^4\theta\,\Big(\phi_\alpha^* e^{2V}\phi_\alpha +
\widetilde\phi_\alpha^* e^{-2V} \widetilde\phi_\alpha\Big),
\end{equation}

\noindent where $V$ is a real gauge superfield; $\phi_\alpha$ and
$\widetilde\phi_\alpha$ are chiral matter superfields. The index
$\alpha$ numerates flavors and goes from 1 to the number of
flavors $N_f$. $e_0$ denotes the bare coupling constant. Also we
will use the notation $\alpha_0 \equiv e_0^2/4\pi$. In the
considered Abelian case the supersymmetric gauge field strength is
given by

\begin{equation}
W_a = \frac{1}{4} \bar D^2 D_a V,
\end{equation}

\noindent where $\bar D_{\dot a}$ and $D_a$ are the left and right
supersymmetric covariant derivatives, respectively.

Calculating quantum corrections we use the background field method
\cite{DeWitt:1965jb,Abbott:1980hw,Abbott:1981ke}. In the Abelian
case, which is considered here, the gauge superfield is presented
as a sum of the quantum field $V$ and the background superfield
$\mbox{\boldmath$V$}$:

\begin{equation}
V \to V_T = V + \mbox{\boldmath$V$}.
\end{equation}

\noindent Moreover, it is convenient to modify the action for the
chiral matter superfields by introducing an auxiliary parameter
$g$ according to the following prescription:

\begin{equation}
e^{2V} \to 1 + g(e^{2V}-1);\qquad e^{-2V} \to 1 + g(e^{-2V}-1),
\end{equation}

\noindent where $V$ is the quantum superfield. The exponents with
the background gauge field remain unchanged. This implies that in
each diagram the degree of $g$ is equal to the number of vertexes
containing at least one line of the quantum gauge superfield.
Introducing the parameter $g$ does not break the background gauge
symmetry, but breaks the quantum gauge symmetry.

In order to regularize the theory we add to its action a term with
higher derivatives and insert the Pauli--Villars determinants into
the generating functional to cancel remaining one-loop divergences
\cite{Faddeev:1980be,Slavnov:1977zf}. The final expression for the
generating functional has the following form
\cite{Stepanyantz:2014ima}:

\begin{eqnarray}\label{Generating_Functional}
&& Z = e^{iW} = \int
DV\,D\phi\,D\widetilde\phi\,\prod\limits_{I=1}^n
\det(V,\mbox{\boldmath$V$},M_I)^{c_I N_f}
\exp\Big(iS_{\mbox{\scriptsize reg}}+ iS_{\mbox{\scriptsize
gf}}+ iS_{\mbox{\scriptsize source}}\Big)\nonumber\\
&& = \int D\mu \exp\Big(i S_{\mbox{\scriptsize total}} + i
S_{\mbox{\scriptsize gf}} + i S_{\mbox{\scriptsize source}} \Big).
\end{eqnarray}

\noindent The masses of the Pauli--Villars fields are proportional
to the parameter $\Lambda$ in the higher derivative terms, $M_I =
a_I \Lambda$, $a_I$ being independent of $e_0$. The coefficients
$c_I = (-1)^{P_I+1}$ should satisfy the conditions

\begin{equation}
\sum\limits_{I=1}^n c_I =1; \qquad \sum\limits_{I=1}^n c_I M_I^2
=0,
\end{equation}

\noindent where $(-1)^{P_I}$ can be considered as the Grassmannian
parity of the corresponding Pauli--Villars fields. Taking into
account that $c_I = \pm 1$, the Pauli--Villars fields can be
treated similarly to the usual fields. Below the usual fields will
correspond to $I=0$, and $M_{I=0}=0$. The action and the gauge
fixing term in (\ref{Generating_Functional}) are

\begin{eqnarray}\label{Actions}
&&\hspace{-3mm} S_{\mbox{\scriptsize total}} =
\frac{1}{4e_0^2}\mbox{Re}\int d^4x\,d^2\theta\,W^a
R(\partial^2/\Lambda^2) W_a + \frac{1}{4e_0^2}\mbox{Re}\int
d^4x\,d^2\theta\,\mbox{\boldmath$W$}^a
\mbox{\boldmath$W$}_a\nonumber\\
&&\hspace{-3mm} + \frac{1}{4} \sum\limits_{I=0}^n
\sum\limits_{\alpha=1}^{N_f} \int d^4x\,d^4\theta\,\Big[
\phi_{\alpha}^* e^{2\mbox{\scriptsize \boldmath$V$}} \Big(1 + g
(e^{2V}-1)\Big) \phi_{\alpha} +\widetilde\phi_{\alpha}^*
e^{-2\mbox{\scriptsize \boldmath$V$}} \Big(1 + g (e^{-2V}-1)\Big)
\widetilde\phi_{\alpha}
\Big]_I\nonumber\\
&&\hspace{-3mm} + \sum\limits_{I=0}^n
\sum\limits_{\alpha=1}^{N_f}\Big(\frac{1}{2}\int
d^4x\,d^2\theta\,M \phi_{\alpha} \widetilde\phi_{\alpha} +
\mbox{c.c.}\Big)_I;\nonumber\\
&&\hspace{-3mm} S_{\mbox{\scriptsize gf}} = - \frac{1}{64 e^2}\int
d^4x\,d^4\theta\, \Big(V R(\partial^{2}/\Lambda^{2}) D^2 \bar D^2
V + V R(\partial^{2}/\Lambda^{2}) \bar D^2 D^2 V\Big),
\end{eqnarray}

\noindent where for the background field strength we use the
notation $\mbox{\boldmath$W$}_a = \bar D^2 D_a
\mbox{\boldmath$V$}/4$ and $e$ is the renormalized coupling
constant. The function $R$ is a sum of 1 corresponding to the
classical action and a higher derivative term. For example, it is
possible to choose $R=1+\partial^{2n}/\Lambda^{2n}$. From Eq.
(\ref{Actions}) we see that the propagator of the gauge superfield
(in the Euclidean space after the Wick rotation) is proportional
to

\begin{equation}
\frac{e_0^2}{R_k k^2} + \frac{(e_0^2 - e^2)}{16 R_k k^4}\Big(\bar
D^2 D^2 + D^2 \bar D^2\Big).
\end{equation}

The source term is constructed by the standard way, but it is also
convenient to include into it sources for the Pauli--Villars
fields. The effective action is defined by

\begin{equation}\label{Gamma}
\Gamma[V,\mbox{\boldmath$V$},\phi_{\alpha
I},\widetilde\phi_{\alpha I}] = W - S_{\mbox{\scriptsize source}},
\end{equation}

\noindent where the sources should be expressed in terms of
fields. Terms in the effective action corresponding to the
two-point functions of the background gauge superfield and the
chiral matter superfields (including the Pauli-Villars ones) can
be presented as\footnote{This expression is written in the
Minkowski space.}

\begin{eqnarray}\label{Two_Point_Gamma}
&& \Gamma^{(2)} - S_{\mbox{\scriptsize gf}} = - \frac{1}{16\pi}
\int
\frac{d^4p}{(2\pi)^4}\,d^4\theta\,\mbox{\boldmath$V$}(\theta,-p)\,\partial^2\Pi_{1/2}
\mbox{\boldmath$V$}(\theta,p)\,
d^{-1}(\alpha_0,\Lambda/p)\nonumber\\
&& + \frac{1}{4} \sum\limits_{I=0}^n
\sum\limits_{\alpha=1}^{N_f}\int \frac{d^4p}{(2\pi)^4}\,
d^4\theta\,\Big(\phi_{\alpha I}^*(\theta,-p) \phi_{\alpha
I}(\theta,p)+ \widetilde\phi_{\alpha I}^*(\theta,-p)
\widetilde\phi_{\alpha I}(\theta,p)\Big)
G_I(\alpha_0,\Lambda/p)\quad\nonumber\\
&& - \Big( \sum\limits_{I=1}^n \sum\limits_{\alpha=1}^{N_f}\int
\frac{d^4p}{(2\pi)^4}\,d^4\theta\, \phi_{\alpha I}(\theta,-p)
\frac{D^2}{16p^2} \widetilde\phi_{\alpha I}(\theta,p)\, M_I
J_I(\alpha_0,\Lambda/p) +\mbox{c.c.}\Big).\qquad
\end{eqnarray}

\noindent Writing explicitly expressions for the functions
entering this equation in the tree approximation we can present
them in the form

\begin{equation}
d^{-1}(\alpha_0,\Lambda/p) = \alpha_0^{-1} + O(1);\qquad
G_I(\alpha_0,\Lambda/p)=1 + O(\alpha_0);\qquad
J_I(\alpha_0,\Lambda/p)=1 + O(\alpha_0).
\end{equation}

\noindent The $\beta$-function defined in terms of the bare
coupling constant can be easily found, if the function
$d^{-1}(\alpha_0,\Lambda/p)$ is known:

\begin{equation}
\frac{d}{d\ln \Lambda}\,
\Big(d^{-1}(\alpha_0,\Lambda/p)-\alpha_0^{-1}\Big)\Big|_{p=0} = -
\frac{d\alpha_0^{-1}(\alpha,\Lambda/\mu)}{d\ln\Lambda} =
\frac{\beta(\alpha_0)}{\alpha_0^2},
\end{equation}

\noindent where $\alpha$ is the renormalized coupling constant,
$\mu$ is the renormalization scale, and the derivative with
respect to $\ln\Lambda$ should be calculated at a fixed value of
the renormalized coupling constant $\alpha$. From this equation we
see that the considered $\beta$-function is completely determined
by the derivative of $d^{-1}(\alpha_0,\Lambda/p)-\alpha_0^{-1}$
with respect to $\ln\Lambda$ in the limit of the vanishing
external momentum, $p\to 0$. This derivative can be found from the
two-point Green function of the background gauge superfield after
the substitution

\begin{equation}
\mbox{\boldmath$V$}(x,\theta) \to  \theta^4 \cdot I(x) \approx
\theta^4,
\end{equation}

\noindent where the regulator $I(x)$ is approximately equal to 1
at finite $x^\mu$ and tends to 0 at the large scale $R \to
\infty$. Introducing the regularized space-time volume

\begin{equation}
{\cal V}_4 \equiv \int d^4x\,I^2 \sim R^4 \to \infty
\end{equation}

\noindent one can easily obtain \cite{Stepanyantz:2014ima}

\begin{equation}\label{Beta_From_Effective_Action}
\frac{1}{2\pi} {\cal V}_4\cdot \frac{d}{d\ln \Lambda}\,
\Big(d^{-1}(\alpha_0,\Lambda/p)-\alpha_0^{-1}\Big)\Big|_{p=0} =
\frac{1}{2\pi} {\cal V}_4\cdot \frac{\beta(\alpha_0)}{\alpha_0^2}
= \frac{d(\Delta\Gamma^{(2)}_{\scriptsize
\mbox{\boldmath$V$}})}{d\ln\Lambda}
\Big|_{\mbox{\boldmath$V$}(x,\theta)=\theta^4},
\end{equation}

\noindent where we use the notation

\begin{equation}
\Delta\Gamma \equiv \Gamma - \frac{1}{4e_0^2}\mbox{Re}\int
d^4x\,d^2\theta\,\mbox{\boldmath$W$}^a \mbox{\boldmath$W$}_a.
\end{equation}

\noindent (The superscript $(2)$ and the subscript
$\mbox{\boldmath$V$}$ extract a part of $\Delta\Gamma$ quadratic
in the background gauge superfield.) The derivative with respect
to $\ln\Lambda$ should be calculated at a fixed value of the
renormalized coupling constant $\alpha$. Below for simplicity we
omit the regulator $I(x)$ in all subsequent equations.

As we already mentioned above, the integrals giving the
$\beta$-function defined in terms of the bare coupling constant
are integrals of double total derivatives in the momentum space
(in these integrals the external momentum vanishes). In Ref.
\cite{Stepanyantz:2014ima} it was shown that this fact is a
consequence of the following identity relating various two-point
Green function of the theory:\footnote{The derivatives of the
effective action, which are written in this equation, coincide
with the corresponding derivatives of the Routhian $\gamma$, which
is used in \cite{Stepanyantz:2014ima}.}

\begin{eqnarray}\label{Double_Total_Derivatives}
&& \frac{d}{d\ln\Lambda} \frac{\partial}{\partial \ln g}
\Big(\frac{1}{2}\int d^8x\,d^8y\,(\theta^4)_x (\theta^4)_y
\frac{\delta^2\Delta\Gamma}{\delta \mbox{\boldmath$V$}_x \delta
\mbox{\boldmath$V$}_y}\Big)\nonumber\\
&& = \frac{i}{4} C(R)_i{}^j \frac{d}{d\ln\Lambda}
\mbox{Tr}\,(\theta^4)_x \Big[y_\mu^*,\Big[y_\mu^*,
\Big(\frac{\delta^2\Gamma}{\delta(\phi_j)_x
\delta(\phi^{*i})_y}\Big)^{-1} + M^{ik}
\Big(\frac{D^2}{8\partial^2}\Big)_x\Big(\frac{\delta^2\Gamma}{\delta
(\phi_k)_x \delta(\phi_j)_y}\Big)^{-1}\qquad\nonumber\\
&& + M^*_{jk} \Big(\frac{\bar
D^2}{8\partial^2}\Big)_x\Big(\frac{\delta^2\Gamma}{\delta
(\phi^{*k})_x \delta(\phi^{*i})_y}\Big)^{-1}
\Big]\Big]_{y=x}-\mbox{singularities}\ =\ -\mbox{singularities},
\end{eqnarray}

\noindent where

\begin{equation}
(y_\mu)^* \equiv x_\mu - i\bar\theta^{\dot a} (\gamma_\mu)_{\dot
a}{}^b \theta_b,
\end{equation}

\noindent and

\begin{equation}
[y_\mu^*,A_{xy}] \equiv (y_\mu^*)_x A_{xy} - A_{xy} (y_\mu^*)_y;
\qquad \mbox{Tr}\, A \equiv \int d^8x\,A_{xx};\qquad \int d^8x
\equiv \int d^4x\,d^4\theta.
\end{equation}

\noindent In the momentum space the commutators with $y_\mu^*$
give derivatives with respect to the loop momentum. That is why
from Eq. (\ref{Double_Total_Derivatives}) we immediately obtain
the statement that the integrals for the $\beta$-function defined
in terms of the bare coupling constant are given by integrals of
double total derivatives. In Eq. (\ref{Double_Total_Derivatives})
we also use the notation

\begin{equation}
\phi_i \equiv (\phi_{\alpha I},\,\widetilde\phi_{\alpha I});\qquad
\phi^{*i} \equiv (\phi_{\alpha I}^*,\,\widetilde \phi_{\alpha
I}^*),\quad i=1,\ldots 2(n+1)N_f.
\end{equation}

\noindent The fields $\phi_i$ include both usual fields and the
Pauli--Villars fields (which have the Grassmannian parity
$(-1)^{P_I}$). In this notation

\begin{equation}
C(R)_i{}^j = \delta_{\alpha\beta}\cdot \delta_{IJ}
\cdot\left(\begin{array}{cc} 1 & 0\\ 0 & 1
\end{array}
\right)
\end{equation}

\noindent and the mass matrix is given by

\begin{equation}
M^{ij} = \delta_{\alpha\beta}\cdot \delta_{IJ} \cdot \left(
\begin{array}{cc}
0 & M_I\\
(-1)^{P_I} M_I & 0
\end{array}
\right),
\end{equation}

\noindent where $2\times 2$ matrix corresponds to the fields
$\phi$ and $\widetilde\phi$. The two-point Green functions of the
chiral matter superfields can be easily found from Eq.
(\ref{Two_Point_Gamma}):

\begin{eqnarray}
&& \frac{\delta^2\Gamma}{\delta(\phi_i)_x \delta (\phi^{*j})_y} =
\delta_{\alpha\beta} \cdot
\delta_{IJ}\cdot \left(\begin{array}{cc} 1 & 0\\
0 & 1
\end{array}\right) G_I
\frac{\bar D_x^2 D_x^2}{16} \delta^8_{xy};\nonumber\\
&& \frac{\delta^2\Gamma}{\delta(\phi_i)_x \delta (\phi_j)_y} = -
\frac{1}{4} \delta_{\alpha\beta} \cdot
\delta_{IJ}\cdot  \left(\begin{array}{cc} 0 & (-1)^{P_I} M_I\\
M_I & 0
\end{array}\right) J_I \bar D_x^2 \delta^8_{xy}.
\end{eqnarray}

\noindent The corresponding inverse functions are constructed,
e.g., in \cite{Stepanyantz:2014ima}:

\begin{eqnarray}\label{Inverse_Green_Functions}
&& \Big(\frac{\delta^2\Gamma}{\delta(\phi_i)_x \delta
(\phi^{*j})_y}\Big)^{-1} = - (-1)^{P_I} \delta_{\alpha\beta} \cdot
\delta_{IJ}\cdot \left(\begin{array}{cc} 1 & 0\\
0 & 1
\end{array}\right)
\frac{G_I \bar D_x^2 D_x^2}{4(\partial^2 G_I^2 + M_I^2 J_I^2)}
\delta^8_{xy};\nonumber\\
&& \Big(\frac{\delta^2\Gamma}{\delta(\phi_i)_x \delta
(\phi_j)_y}\Big)^{-1} = - \delta_{\alpha\beta} \cdot
\delta_{IJ}\cdot  \left(\begin{array}{cc} 0 & M_I\\
(-1)^{P_I} M_I & 0
\end{array}\right)
\frac{J_I \bar D_x^2}{\partial^2 G_I^2 + M_I^2 J_I^2}
\delta^8_{xy}.
\end{eqnarray}

\noindent The term "singularities" in Eq.
(\ref{Double_Total_Derivatives}) denotes the sum of singular
contributions, which appear due to the identity

\begin{equation}\label{Delta_Commutator}
[x^\mu,\frac{\partial_\mu}{\partial^4}] =
[-i\frac{\partial}{\partial p_\mu}, -\frac{ip_\mu}{p^4}] = -2\pi^2
\delta^4(p_E) = -2\pi^2 i \delta^4(p) = -2\pi^2
i\delta^4(\partial).
\end{equation}

\noindent According to Refs.
\cite{Stepanyantz:2011jy,Stepanyantz:2014ima} the sum of these
singular contributions gives the NSVZ $\beta$ relation for the RG
functions defined in terms of the bare coupling constant in all
orders of the perturbation theory (in the case of using the
considered version of the higher derivative regularization).

\section{Three-loop verification of the relation between the two-point
Green functions}
\label{Section_Explicit_Three_Loop}

\subsection{Double total derivatives in the three-loop approximation}
\hspace{\parindent}\label{Subsection_Double_Total_Derivatives}

\begin{figure}[h]
\begin{picture}(0,2.0)
\put(0.0,1.3){$(1)$} \put(0.7,-0.8){$g^2 N_f$}
\put(0.4,0){\includegraphics[scale=0.43]{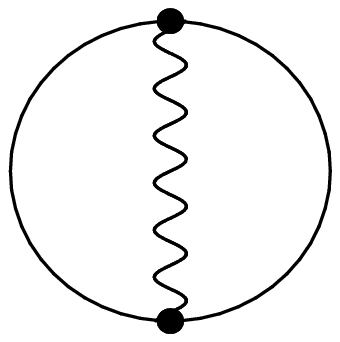}}
\put(2.35,1.3){$(2)$} \put(3.05,-0.8){$g^3 N_f$}
\put(2.75,0){\includegraphics[scale=0.43]{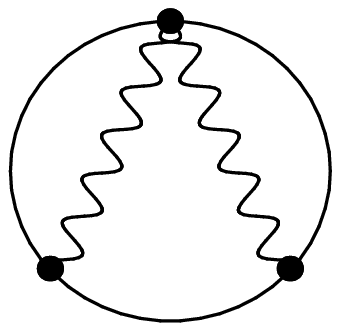}}
\put(4.7,1.3){$(3)$} \put(5.4,-0.8){$g^3 N_f^2$}
\put(5.1,0){\includegraphics[scale=0.43]{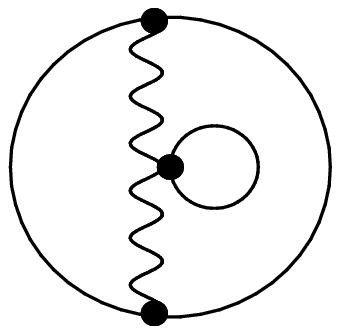}}
\put(7.05,1.3){$(4)$} \put(7.75,-0.8){$g^4 N_f$}
\put(7.45,0){\includegraphics[scale=0.43]{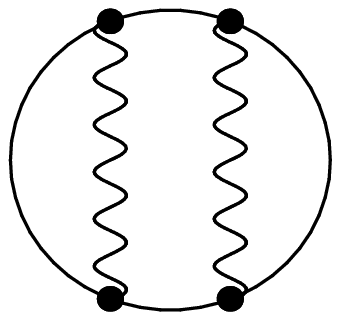}}
\put(9.4,1.3){$(5)$} \put(10.2,-0.8){$g^4 N_f$}
\put(9.8,0){\includegraphics[scale=0.43]{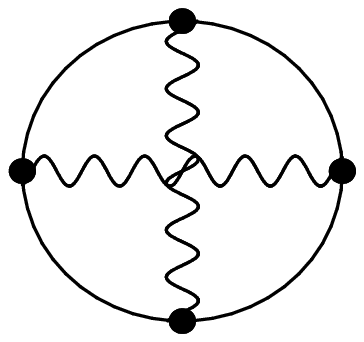}}
\put(11.75,1.3){$(6)$} \put(12.55,-0.8){$g^4 N_f^2$}
\put(12.15,0){\includegraphics[scale=0.43]{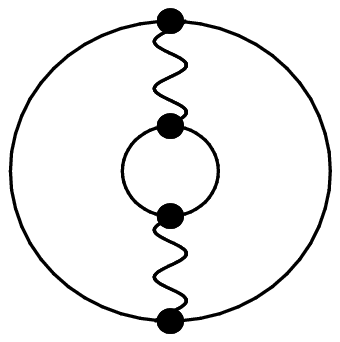}}
\put(14.05,1.3){$(7)$} \put(14.83,-0.8){$g N_f$}
\put(14.35,0){\includegraphics[scale=0.43]{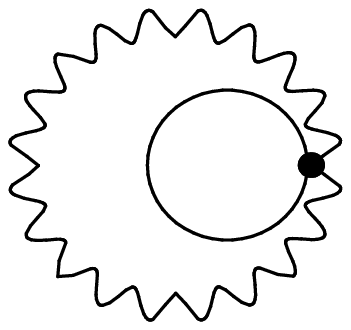}}

\put(0.1,-2.2){\includegraphics[scale=0.25]{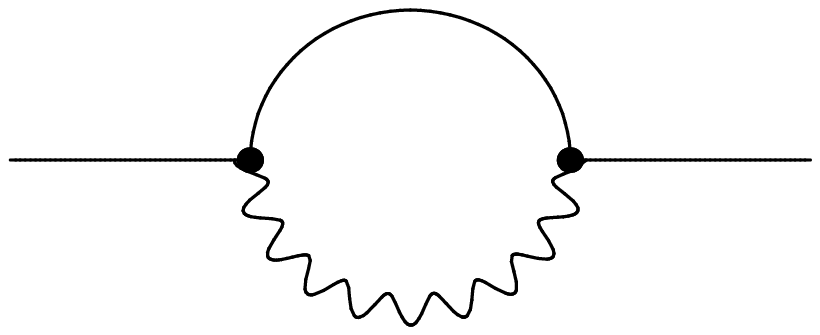}}

\put(2.45,-2.2){\includegraphics[scale=0.25]{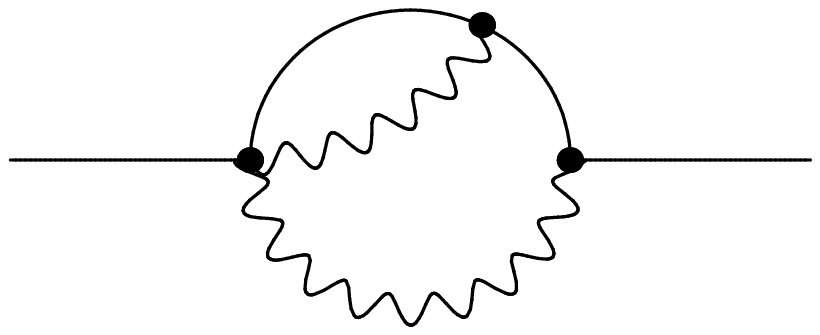}}
\put(2.35,-3.3){\includegraphics[scale=0.2]{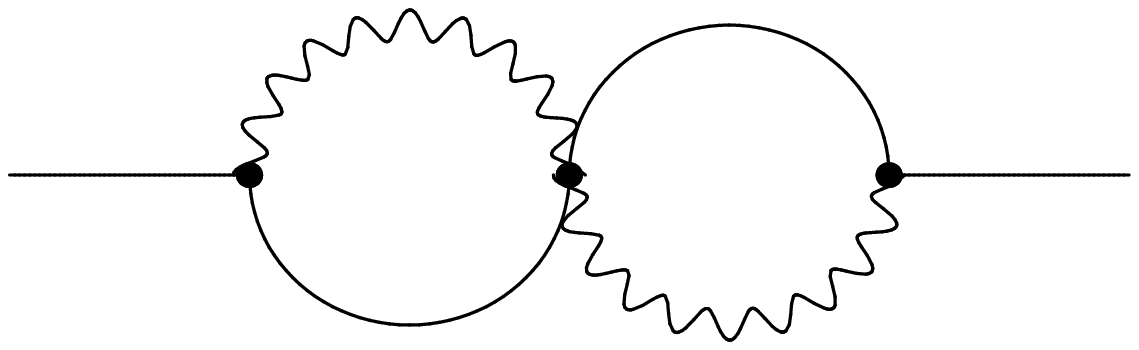}}

\put(5.05,-2.2){\includegraphics[scale=0.23]{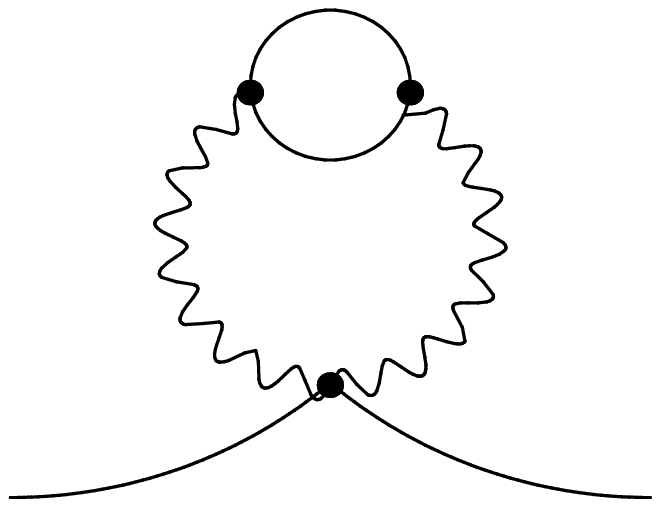}}
\put(4.9,-3.3){\includegraphics[scale=0.23]{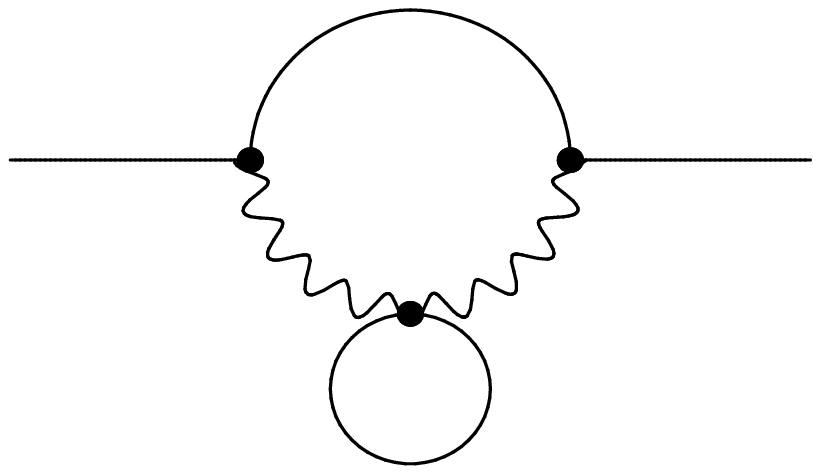}}

\put(7.25,-2.2){\includegraphics[scale=0.25]{dg_gamma1.eps}}
\put(7.25,-3.3){\includegraphics[scale=0.25]{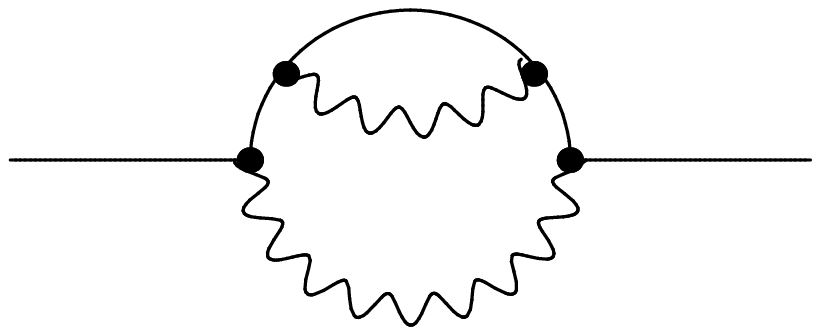}}

\put(9.6,-2.2){\includegraphics[scale=0.25]{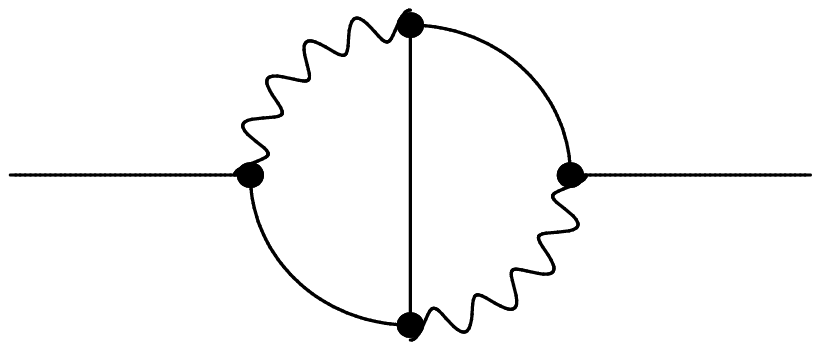}}

\put(11.85,-2.3){\includegraphics[scale=0.25]{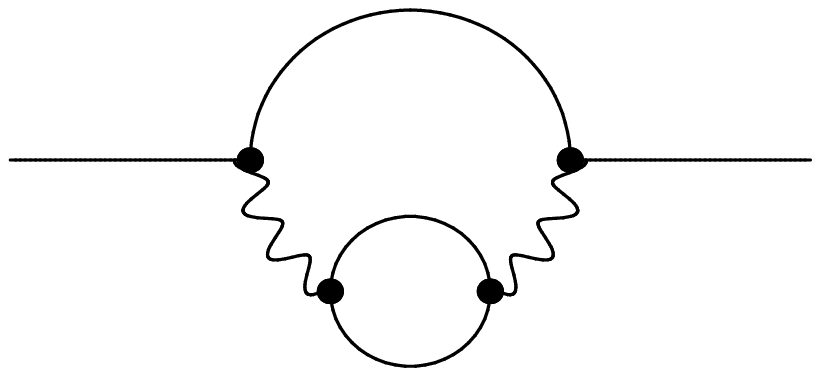}}

\put(14.1,-2.4){\includegraphics[scale=0.233]{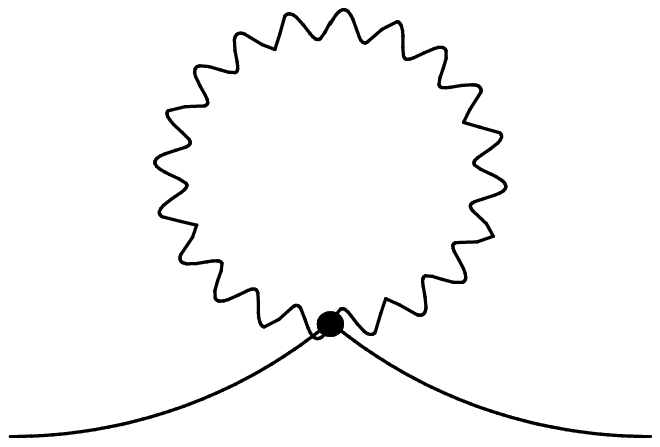}}

\end{picture}
\vspace*{3.3cm} \caption{Two- and three-loop graphs. Diagrams
which contribute to the $\beta$-function can be obtained from them
by attaching two external lines of the gauge superfield. Below we
present factors corresponding to each of these diagrams and the
corresponding diagrams contributing to the two-point Green
function of the matter superfield.}\label{Figure_Beta_Diagrams}
\end{figure}

\begin{figure}[h]
\begin{picture}(0,1.2)
\put(0.4,-1.1){\includegraphics[scale=0.43]{3loop1.eps}}
\put(2.2,-0.41){$\vector(1,0){1}$}
\put(3.6,-0.5){$\left\{\vphantom{\begin{array}{c} 1\\ 1\\
1\\ 1\\ 1\\ 1
\end{array}} \right.$}
\put(4.4,0){\includegraphics[scale=0.35]{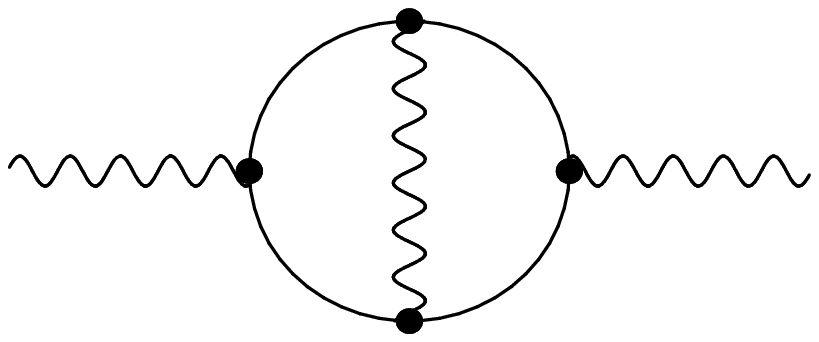}}
\put(8.4,0){\includegraphics[scale=0.35]{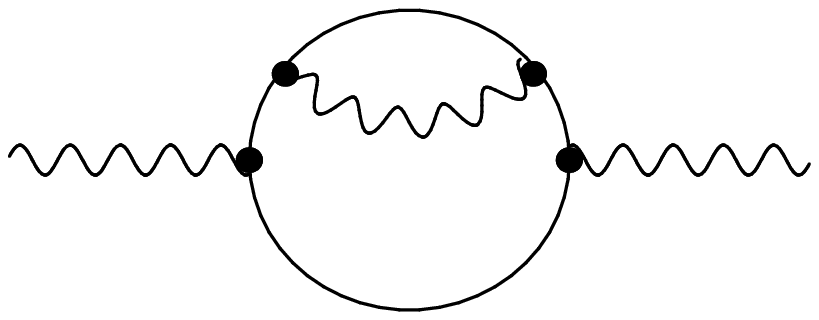}}
\put(12.4,0){\includegraphics[scale=0.35]{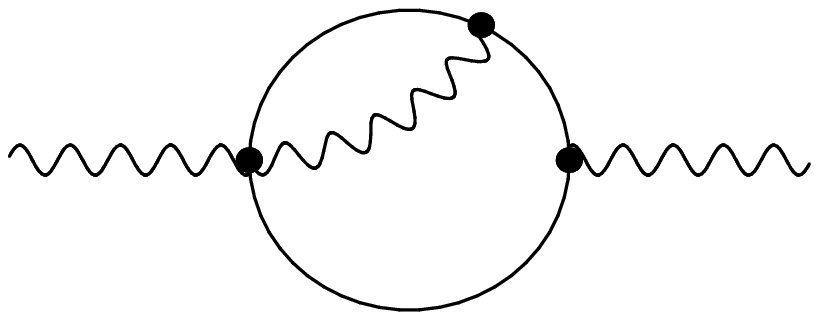}}
\put(4.4,-2){\includegraphics[scale=0.35]{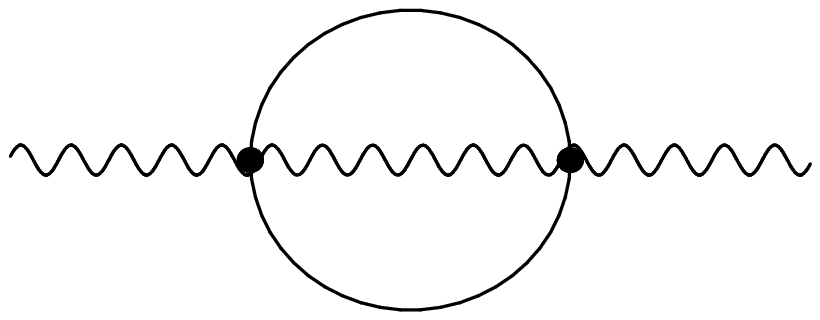}}
\put(8.7,-2){\includegraphics[scale=0.35]{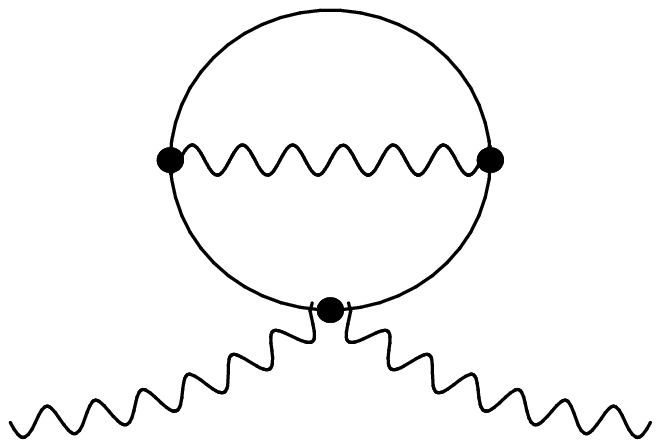}}
\put(12.75,-2){\includegraphics[scale=0.35]{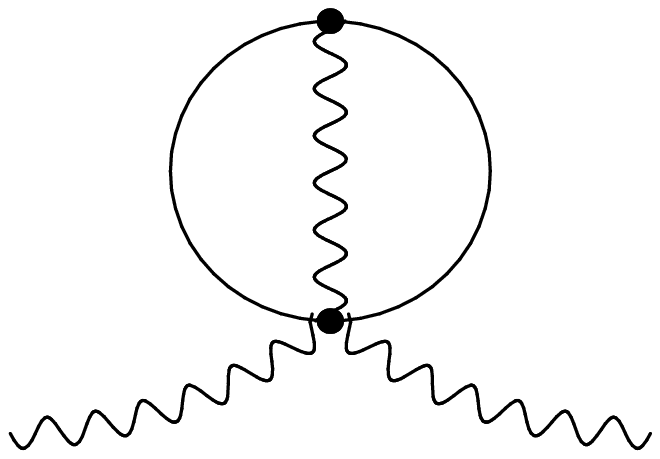}}
\end{picture}
\vspace*{2cm} \caption{Diagrams giving the two-point Green
function of the background gauge superfield in the two-loop
approximation. They correspond to the diagram (1) in Fig.
\ref{Figure_Beta_Diagrams}.}\label{Figure_Two_Loop}
\end{figure}

In order to verify Eq. (\ref{Double_Total_Derivatives}) it is
necessary to calculate the two-point Green function of the
background gauge superfield assuming that $g\ne 1$. In the
three-loop approximation this Green function is given by the sum
of diagrams presented in Fig. \ref{Figure_Beta_Diagrams} to which
one should attach two external lines of the background gauge
superfield.\footnote{The one-loop graph is not included, because
the one-loop approximation in this approach should be considered
separately.} As an example, in Fig. \ref{Figure_Two_Loop} we
present two-loop diagrams which correspond to the two-loop diagram
(1) in Fig. \ref{Figure_Beta_Diagrams}. Each vertex containing the
line of the quantum gauge superfield gives the factor $g$, and
each closed loop of the matter superfield gives the factor $N_f$.
The overall factors for all diagrams are also written in Fig.
\ref{Figure_Beta_Diagrams}. Also in this figure below each graph
we present the corresponding diagrams contributing to the
two-point Green functions of the chiral matter superfields. These
diagrams are obtained by cutting matter lines in the considered
graph.

For the case $N_f=1$, $g=1$ all diagrams giving the three-loop
$\beta$-function were calculated in Ref. \cite{Soloshenko:2003nc}.
Using the results of Ref. \cite{Soloshenko:2003nc} one can find
expressions for all diagrams in Fig. \ref{Figure_Beta_Diagrams}.
As we discussed above, it is convenient to substitute the
background gauge superfield corresponding to the external lines
with $\theta^4$. Below we collect contributions of all diagrams to
the derivative of the expression
(\ref{Beta_From_Effective_Action}) with respect to $\ln g$ in the
Euclidean space (after the Wick rotation). Note that all terms
proportional to $e_0^6$, which are not essential in the considered
approximation, are omitted. The result has the following form:

\begin{eqnarray}\label{Diagram1}
&& \frac{\partial}{\partial\ln g} \frac{d}{d\ln\Lambda}\,
\mbox{diagram $(1)$} = {\cal V}_4 \cdot 4 g^2 N_f
\frac{d}{d\ln\Lambda} \int \frac{d^4q}{(2\pi)^4}
\frac{d^4k}{(2\pi)^4} \frac{1}{k^2 R_k}
\nonumber\\
&&\qquad\qquad\qquad \times \frac{\partial}{\partial q^\mu}
\frac{\partial}{\partial q_\mu} \sum\limits_{I=0}^n (-1)^{P_I}
\Bigg(\frac{e_0^2}{(q^2+M_I^2)\left((q+k)^2 + M_I^2\right)}-
\frac{2(e_0^2-e^2)}{k^2 (q^2 + M_I^2)}\Bigg);\qquad
\end{eqnarray}

\begin{eqnarray}\label{Diagram2}
&& \frac{\partial}{\partial\ln g} \frac{d}{d\ln\Lambda}\,
\mbox{diagram $(2)$} = - {\cal V}_4 \cdot 24 g^3 N_f
\frac{d}{d\ln\Lambda} \int \frac{d^4q}{(2\pi)^4}
\frac{d^4k}{(2\pi)^4} \frac{d^4l}{(2\pi)^4} \frac{e_0^4}{k^2 R_k
l^2 R_l} \nonumber\\
&&\qquad\qquad\qquad\quad \times \frac{\partial}{\partial q^\mu}
\frac{\partial}{\partial q_\mu} \sum\limits_{I=0}^n
(-1)^{P_I}\frac{1}{(q^2+M_I^2)\left((q+k)^2 +
M_I^2\right)\left((q+l)^2 + M_I^2\right)};\qquad
\end{eqnarray}

\begin{eqnarray}\label{Diagram3}
&& \frac{\partial}{\partial\ln g} \frac{d}{d\ln\Lambda}\,
\mbox{diagram $(3)$} = - {\cal V}_4 \cdot 24 g^3 N_f^2
\frac{d}{d\ln\Lambda} \int \frac{d^4q}{(2\pi)^4}
\frac{d^4k}{(2\pi)^4} \frac{d^4l}{(2\pi)^4} \frac{e_0^4}{k^4
R_k^2} \nonumber\\
&&\qquad \times \Big( \frac{\partial}{\partial q^\mu}
\frac{\partial}{\partial q_\mu} + \frac{\partial}{\partial l^\mu}
\frac{\partial}{\partial l_\mu}\Big) \sum\limits_{I,J=0}^n
(-1)^{P_I+P_J} \frac{1}{(l^2 + M_I^2)(q^2+M_J^2)\left((q+k)^2 +
M_J^2\right)};\qquad
\end{eqnarray}

\begin{eqnarray}\label{Diagram4}
&& \frac{\partial}{\partial\ln g} \frac{d}{d\ln\Lambda}\,
\mbox{diagram $(4)$} = {\cal V}_4 \cdot 16 g^4 N_f
\frac{d}{d\ln\Lambda} \int \frac{d^4q}{(2\pi)^4}
\frac{d^4k}{(2\pi)^4} \frac{d^4l}{(2\pi)^4} \frac{e_0^4}{k^2 R_k
l^2 R_l} \nonumber\\
&&\qquad\qquad\qquad\ \ \times \frac{\partial}{\partial q^\mu}
\frac{\partial}{\partial q_\mu} \sum\limits_{I=0}^n (-1)^{P_I}
\frac{q^2 - M_I^2}{(q^2+M_I^2)^2\left((q+k)^2 +
M_I^2\right)\left((q+l)^2 + M_I^2\right)};\qquad
\end{eqnarray}

\begin{eqnarray}\label{Diagram5}
&& \frac{\partial}{\partial\ln g} \frac{d}{d\ln\Lambda}\,
\mbox{diagram $(5)$} = {\cal V}_4 \cdot 8 g^4 N_f
\frac{d}{d\ln\Lambda} \int \frac{d^4q}{(2\pi)^4}
\frac{d^4k}{(2\pi)^4} \frac{d^4l}{(2\pi)^4} \frac{e_0^4}{k^2 R_k
l^2 R_l} \frac{\partial}{\partial q^\mu} \frac{\partial}{\partial
q_\mu} \ \nonumber\\
&&\qquad \times \sum\limits_{I=0}^n (-1)^{P_I} \frac{(k+l+2q)^2 +
2 M_I^2}{(q^2+M_I^2)\left((q+k)^2 + M_I^2\right)\left((q+l)^2 +
M_I^2\right)\left((q+k+l)^2 + M_I^2\right)};\qquad
\end{eqnarray}

\begin{eqnarray}\label{Diagram6}
&& \frac{\partial}{\partial\ln g} \frac{d}{d\ln\Lambda}\,
\mbox{diagram $(6)$} = -16 {\cal V}_4 \cdot g^4 N_f^2
\frac{d}{d\ln\Lambda} \int \frac{d^4q}{(2\pi)^4}
\frac{d^4k}{(2\pi)^4} \frac{d^4l}{(2\pi)^4} \frac{e_0^4}{k^4
R_k^2} \nonumber\\
&&\quad \times \frac{\partial}{\partial q^\mu}
\frac{\partial}{\partial q_\mu} \sum\limits_{I,J=0}^n
(-1)^{P_I+P_J} \frac{k^2 - (k+q)^2 - q^2 -(l+k)^2 - l^2 - 2M_I^2
-2M_J^2}{(q^2+M_I^2) \left((q+k)^2 +
M_I^2\right)(l^2+M_J^2)\left((l+k)^2 + M_J^2\right)}. \qquad
\end{eqnarray}

\begin{eqnarray}\label{Diagram7}
&& \frac{\partial}{\partial\ln g} \frac{d}{d\ln\Lambda}\,
\mbox{diagram $(7)$} = 4 {\cal V}_4 \cdot g N_f
\frac{d}{d\ln\Lambda} \int \frac{d^4q}{(2\pi)^4}
\frac{d^4k}{(2\pi)^4} \frac{(e_0^2 - e^2)}{k^4
R_k} \nonumber\\
&&\qquad\qquad\qquad\qquad\qquad\qquad\qquad\qquad\qquad\qquad
\quad\ \ \times \frac{\partial}{\partial q^\mu}
\frac{\partial}{\partial q_\mu} \sum\limits_{I=0}^n (-1)^{P_I}
\frac{1}{\left(q^2 + M_I^2\right)}. \qquad
\end{eqnarray}

\noindent We see that each graph is given by an integral of double
total derivatives, as it was argued in \cite{Smilga:2004zr}.

\subsection{Terms proportional to $e_0^2$ in Eq. (\ref{Double_Total_Derivatives})}
\label{Subsection_Two_Loop}\hspace{\parindent}

From the above equations we see that expressions for all graphs in
Fig. \ref{Figure_Beta_Diagrams} are integrals of double total
derivatives. Eq. (\ref{Double_Total_Derivatives}) allows to
construct these integrals of double total derivatives without
calculating the three-loop diagrams. In order to use this equation
one should calculate the two-point Green functions of the chiral
matter superfields in the previous (two-loop) approximation. Let
us first demonstrate this for the first (two-loop) diagram in Fig.
\ref{Figure_Beta_Diagrams}. Cutting the line of the matter
superfield we obtain the corresponding (one-loop) diagram for the
two-point Green functions of the matter superfields, which is also
presented in Fig. \ref{Figure_Beta_Diagrams}. Calculating this
diagram we obtain its contributions to the functions $G_I$ and
$J_I$ (in the Euclidean space after the Wick rotation, as
functions of the Euclidean momentum $q_\mu$):

\begin{eqnarray}\label{G1}
&& \Delta G^{(1)}_I = - g^2 \int \frac{d^4k}{(2\pi)^4}\frac{2}{k^2
R_k} \Bigg(\frac{e_0^2}{\left((q+k)^2 + M_I^2\right)}
- \frac{(e_0^2- e^2)\left((q+k)^2 + q^2\right)}{k^2 \left((q+k)^2 + M_I^2\right)}\Bigg);\\
\label{J1} && \Delta J^{(1)}_I= g^2 \int
\frac{d^4k}{(2\pi)^4}\frac{4 (e_0^2- e^2) q^2}{k^4 R_k
\left((q+k)^2 + M_I^2\right)}.
\end{eqnarray}

\noindent The difference $e_0^2- e^2$ is evidently proportional to
$e_0^4$. Therefore, the contributions proportional to $e_0^2-e^2$
are essential only in the next order. This implies that in the
one-loop approximation

\begin{eqnarray}
&& G_I(\alpha_0,\Lambda/q)=1 -  \int \frac{d^4k}{(2\pi)^4}\frac{2
e_0^2 g^2}{k^2 R_k \left((q+k)^2 +
M_I^2\right)} + O(e_0^4);\nonumber\\
&& J_I(\alpha_0,\Lambda/q) = 1 + O(e_0^4).\vphantom{\Big(}
\end{eqnarray}

\noindent Then we construct the inverse Green functions entering
Eq. (\ref{Double_Total_Derivatives}) using Eq.
(\ref{Inverse_Green_Functions}). In the momentum representation
they are proportional to

\begin{eqnarray}\label{Inverse_Two_Loop}
&& \Big(\frac{\delta^2\Gamma}{\delta(\phi_i)_x \delta
(\phi^{*j})_y}\Big)^{-1} \sim \frac{G_I}{4(q^2 G_I^2 + M_I^2
J_I^2)}\nonumber\\
&&\qquad\qquad\qquad = \frac{1}{4(q^2 + M_I^2)}\Big(1 + \frac{q^2
- M_I^2}{q^2 + M_I^2} \int \frac{d^4k}{(2\pi)^4}\frac{2 e_0^2
g^2}{k^2 R_k \left((q+k)^2 + M_I^2\right)} +
O(e_0^4)\Big);\quad\nonumber\\
&& \Big(\frac{\delta^2\Gamma}{\delta(\phi_i)_x \delta
(\phi_j)_y}\Big)^{-1} \sim \frac{ M_I J_I}{q^2 G_I^2 + M_I^2
J_I^2}\nonumber\\
&&\qquad\qquad\qquad = \frac{M_I}{q^2 + M_I^2}\Big(1 +
\frac{q^2}{q^2 + M_I^2} \int \frac{d^4k}{(2\pi)^4}\frac{4 e_0^2
g^2}{k^2 R_k \left((q+k)^2 + M_I^2\right)} + O(e_0^4)\Big).\qquad
\end{eqnarray}

\noindent These functions are needed for calculating the
expression\footnote{The integral over $d^4\theta$ comes from
$\mbox{Tr}$ in Eq. (\ref{Double_Total_Derivatives}).}

\begin{eqnarray}
&&\hspace*{-5mm} \frac{i}{4} C(R)_i{}^j \int d^4\theta_x
(\theta^4)_x \Bigg(\Big(\frac{\delta^2\Gamma}{\delta(\phi_j)_x
\delta(\phi^{*i})_y}\Big)^{-1} + M^{ik}
\Big(\frac{D^2}{8\partial^2}\Big)_x\Big(\frac{\delta^2\Gamma}{\delta
(\phi_k)_x \delta(\phi_j)_y}\Big)^{-1} + M^*_{jk} \Big(\frac{\bar
D^2}{8\partial^2}\Big)_x\nonumber\\
&&\hspace*{-5mm} \times \Big(\frac{\delta^2\Gamma}{\delta
(\phi^{*k})_x \delta(\phi^{*i})_y}\Big)^{-1}\Bigg)\Bigg|_{\theta_y
= \theta_x} = 2 N_f \sum\limits_{I=0}^n (-1)^{P_I} \int
\frac{d^4q}{(2\pi)^4} \Bigg(\Big(\frac{1}{q^2 +
M_I^2} + \frac{M_I^2}{q^2 (q^2 + M_I^2)}\Big)\nonumber\\
&&\hspace*{-5mm} + \Big(\frac{q^2- M_I^2}{(q^2 + M_I^2)^2} +
\frac{2M_I^2}{(q^2 + M_I^2)^2} \Big) \int
\frac{d^4k}{(2\pi)^4}\frac{2 e_0^2 g^2}{k^2 R_k \left((q+k)^2 +
M_I^2\right)} +
O(e_0^4)\Bigg) \exp\Big(iq_\mu(x^\mu-y^\mu)\Big)\nonumber\\
&&\hspace*{-5mm} = 4 N_f \sum\limits_{I=0}^n (-1)^{P_I} \int
\frac{d^4q}{(2\pi)^4} \frac{1}{q^2 + M_I^2} \int
\frac{d^4k}{(2\pi)^4}\frac{e_0^2 g^2}{k^2 R_k \left((q+k)^2 +
M_I^2\right)}  \exp\Big(iq_\mu(x^\mu-y^\mu)\Big)\nonumber\\
&&\hspace*{-5mm} + O(e_0^4),\vphantom{\frac{1}{2}}
\end{eqnarray}

\noindent where $x^\mu$ and $y^\mu$ denote Euclidean coordinates.
Deriving the last equality we take into account that one of the
terms is proportional to

\begin{equation}
\sum\limits_{I=0}^n (-1)^{P_I} = 1 - \sum\limits_{I=1}^n c_I = 0.
\end{equation}

\noindent At the next step it is necessary to calculate the
commutators with $y_\mu^*$. Taking into account that for
$\theta$-s trace of the commutator always vanishes, we can replace
commutators with $y_\mu^*$ by the commutators with $x_\mu$. These
commutators in the momentum representation give

\begin{equation}
[x_\mu,[x^\mu, \ldots]]_{M} = - [x_\mu,[x^\mu, \ldots]]_{E}  \to
\frac{\partial}{\partial q^\mu} \frac{\partial}{\partial q_\mu},
\end{equation}

\noindent which acts on the exponent. Then we integrate by parts
and calculate the remaining part of the trace by setting $y=x$ and
integrating over $d^4x$. This gives

\begin{eqnarray}\label{Two_Loop_Result}
&& \frac{d}{d\ln\Lambda} \frac{\partial}{\partial \ln g}
\Big(\frac{1}{2}\int d^8x\,d^8y\,(\theta^4)_x (\theta^4)_y
\frac{\delta^2\Delta\Gamma}{\delta \mbox{\boldmath$V$}_x \delta
\mbox{\boldmath$V$}_y}\Big) = 4 {\cal V}_4 N_f \sum\limits_{I=0}^n
(-1)^{P_I} \frac{d}{d\ln\Lambda}\nonumber\\
&& \times \int \frac{d^4q}{(2\pi)^4} \frac{\partial}{\partial
q^\mu} \frac{\partial}{\partial q_\mu} \Bigg(\frac{1}{(q^2 +
M_I^2)} \int \frac{d^4k}{(2\pi)^4}\frac{e_0^2 g^2}{k^2 R_k
\left((q+k)^2 + M_I^2\right)}\Bigg) + O(e_0^4).\qquad
\end{eqnarray}

\noindent (Certainly, the presence of the regulator $I^2(x)$ is
assumed, so that the integration over $d^4x$ gives ${\cal V}_4\to
\infty$.) The expression in the right hand side of this equation
coincides with the result of explicit two-loop calculations, which
is given by the first term in Eq. (\ref{Diagram1}). Thus, Eq.
(\ref{Double_Total_Derivatives}) has been verified in the two-loop
approximation.

The integrand in Eq. (\ref{Two_Loop_Result}) contains
singularities due to the identity

\begin{equation}
\frac{\partial}{\partial q^\mu} \frac{\partial}{\partial q_\mu}
\frac{1}{q^2} = -4\pi^2 \delta^4(q).
\end{equation}

\noindent If we define the integrals by the standard way, then we
should surround the singularities by small spheres and take into
account the corresponding surface integrals. As a consequence, the
integral in Eq. (\ref{Two_Loop_Result}) corresponds to the trace
of commutator {\it minus singularities}. This can be illustrated
by the following simple example. If $f(q^2)$ is a noningular
function which rapidly descreases at the infinity, then the
vanishing trace of commutator corresponds, e.g., to the object

\begin{eqnarray}\label{Example}
&& 0 = \int \frac{d^4q}{(2\pi)^4}
\frac{\mbox{\boldmath$\partial$}}{\mbox{\boldmath$\partial$}
q^\mu} \Big(\frac{q^\mu}{q^4} f(q^2)\Big) \equiv \int
\frac{d^4q}{(2\pi)^4} \Big(\frac{\partial}{\partial q^\mu}
\frac{q^\mu}{q^4} \cdot f(q^2) + \frac{q^\mu}{q^4} \frac{\partial
f(q^2)}{\partial q^\mu}\Big)\nonumber\\
&& = \int \frac{d^4q}{(2\pi)^4} \Big(2\pi^2 \delta^4(q) f(q^2) +
\frac{q^\mu}{q^4} \frac{\partial f(q^2)}{\partial q^\mu}\Big) =
\frac{1}{8\pi^2} f(0) + \int \frac{d^4q}{(2\pi)^4}
\frac{\partial}{\partial q^\mu} \Big(\frac{q^\mu}{q^4}
f(q^2)\Big),\qquad
\end{eqnarray}

\noindent where the last integral is defined by the standard way.

Let us integrate the equality (\ref{Two_Loop_Result}) over $\ln g$
from $g=0$ to $g=1$. For $g=1$ the theory coincides with ${\cal
N}=1$ SQED with $N_f$ flavors. The case $g=0$ corresponds to the
theory without quantum gauge superfield. Therefore, the two-point
Green function of the background gauge superfield in this case is
contributed only by the one-loop diagram. Thus, using Eq.
(\ref{Beta_From_Effective_Action}) we obtain

\begin{eqnarray}
&& \frac{\beta(\alpha_0)}{\alpha_0^2} -
\frac{\beta_{\mbox{\scriptsize 1-loop}}(\alpha_0)}{\alpha_0^2} =
4\pi N_f  \frac{d}{d\ln\Lambda} \int \frac{d^4k}{(2\pi)^4}
\frac{d^4q}{(2\pi)^4} \frac{e_0^2}{k^2 R_k}
\frac{\partial}{\partial q^\mu} \frac{\partial}{\partial
q_\mu}\nonumber\\
&&\qquad\qquad\qquad\qquad \times \Bigg(\frac{1}{q^2 (q+k)^2} -
\sum\limits_{I=1}^n c_I \frac{1}{(q^2 + M_I^2) \left((q+k)^2 +
M_I^2\right)}\Bigg) + O(e_0^4).\qquad
\end{eqnarray}

\noindent This integral does not vanish due to the singularities
of the integrand. Surrounding singular points $q^\mu=0$ and $q^\mu
= -k^\mu$ by small spheres and calculating integrals of total
derivatives taking into account all surface integrals we easily
find

\begin{equation}
\frac{\beta(\alpha_0)}{\alpha_0^2} =
\frac{\beta_{\mbox{\scriptsize 1-loop}}(\alpha_0)}{\alpha_0^2}
+\frac{2}{\pi} N_f  \frac{d}{d\ln\Lambda} \int
\frac{d^4k}{(2\pi)^4} \frac{e_0^2}{k^4 R_k} + O(e_0^4) =
\frac{\beta_{\mbox{\scriptsize 1-loop}}(\alpha_0)}{\alpha_0^2} +
\frac{\alpha_0 N_f}{\pi^2} + O(\alpha_0^2).
\end{equation}

\noindent Substituting the expression for the one-loop
$\beta$-function of the considered theory we reobtain the
well-known result (see, e.g., \cite{Jones:1977zr,Shifman:1985tj})
for the two-loop $\beta$-function

\begin{equation}
\beta(\alpha_0) = \frac{N_f \alpha_0^2}{\pi}\Big(1 +
\frac{\alpha_0}{\pi} + O(\alpha_0^2)\Big).
\end{equation}

\subsection{Terms proportional to $e_0^4$ in Eq. (\ref{Double_Total_Derivatives})}
\label{Subsection_Two_Loop}\hspace{\parindent}

Verification of Eq. (\ref{Double_Total_Derivatives}) in the next
order is made similarly. However, in the three-loop approximation
one encounters some subtleties, which are absent in the previous
order. Let us sequentially discuss application of Eq.
(\ref{Double_Total_Derivatives}) to finding expressions for the
three-loop diagrams presented in Fig. \ref{Figure_Beta_Diagrams}.

1. We will start with the graph (1) in Fig.
\ref{Figure_Beta_Diagrams}. The corresponding contributions to the
functions $G_I$ and $J_I$ are given by Eqs. (\ref{G1}) and
(\ref{J1}). Terms which are essential in the considered
approximation are proportional to $e_0^2-e^2$:

\begin{eqnarray}
&& \Delta G^{(1)}_I \leftarrow g^2 \int
\frac{d^4k}{(2\pi)^4}\frac{2
(e_0^2- e^2)\left((q+k)^2 + q^2\right)}{k^4 R_k \left((q+k)^2 + M_I^2\right)};\nonumber\\
&& \Delta J^{(1)}_I \leftarrow g^2 \int
\frac{d^4k}{(2\pi)^4}\frac{4 (e_0^2- e^2) q^2}{k^4 R_k
\left((q+k)^2 + M_I^2\right)}.
\end{eqnarray}

\noindent Repeating the calculation similarly to the one made in
the previous section, we obtain

\begin{eqnarray}
&& \frac{d}{d\ln\Lambda} \frac{\partial}{\partial \ln g}
\Big(\frac{1}{2}\int d^8x\,d^8y\,(\theta^4)_x (\theta^4)_y
\frac{\delta^2\Delta\Gamma}{\delta \mbox{\boldmath$V$}_x \delta
\mbox{\boldmath$V$}_y}\Big) \leftarrow - 4 {\cal V}_4 \cdot g^2
N_f \frac{d}{d\ln\Lambda} \nonumber\\
&&\qquad\qquad\quad \times \int \frac{d^4q}{(2\pi)^4}
\frac{d^4k}{(2\pi)^4} \frac{(e_0^2 - e^2)}{k^4 R_k^2}
\frac{\partial}{\partial q^\mu} \frac{\partial}{\partial q_\mu}
\sum\limits_{I=0}^n (-1)^{P_I} \frac{(k+q)^2 + q^2 +
2M_I^2}{(q^2+M_I^2)\left((q+k)^2 + M_I^2\right)}.\qquad\quad
\end{eqnarray}

\noindent This contribution coincides with the second term in Eq.
(\ref{Diagram1}) after some simple transformations. (The first
term was obtained earlier in the previous section.)

2. Diagrams corresponding to the second graph in Fig.
\ref{Figure_Beta_Diagrams} give the following contributions to the
functions $G_I$ and $J_I$:

\begin{eqnarray}
&& \Delta G^{(2)}_I = g^3 \int \frac{d^4k}{(2\pi)^4}
\frac{d^4l}{(2\pi)^4} \frac{4 e_0^4}{k^2 R_k l^2 R_l}
\Bigg(\frac{1}{\left((q+k)^2 +
M_I^2\right)\left((q+l)^2 + M_I^2\right)}\nonumber\\
&& \qquad\qquad\qquad\qquad\qquad\qquad\qquad\qquad\qquad +
\frac{2}{\left((q+k)^2 + M_I^2\right)\left((q+k+l)^2 +
M_I^2\right)}
\Bigg);\qquad \nonumber\\
&& \Delta J^{(2)}_I=0.
\end{eqnarray}

\noindent Repeating the above calculations in this case we obtain
the contribution (\ref{Diagram2}).

3. Diagram (3) in Fig. \ref{Figure_Beta_Diagrams} contains two
closed loops of the matter superfields. Constructing the
corresponding diagrams which contribute to the two-point function
of the matter superfields it is necessary to take into account a
possibility of cutting each of these loops. The diagrams obtained
as a result of this procedure are presented in the third column of
Fig. \ref{Figure_Beta_Diagrams}. They give

\begin{eqnarray}
&& \Delta G^{(3)}_I = g^3 N_f \int \frac{d^4k}{(2\pi)^4}
\frac{d^4l}{(2\pi)^4} \frac{4 e_0^4}{k^4 R_k^2}
\Bigg(\frac{2}{\left((q+k)^2 +
M_I^2\right)}\sum\limits_{J=0}^n (-1)^{P_J} \frac{1}{l^2 + M_J^2} \nonumber\\
&&\qquad\qquad\qquad\qquad\qquad\qquad\qquad\qquad\qquad +
\sum\limits_{J=0}^n (-1)^{P_J} \frac{1}{\left(l^2 + M_J^2\right)
\left((l+k)^2 + M_J^2\right)}
\Bigg);\qquad \nonumber\\
&& \Delta J^{(3)}_I=0.\vphantom{\frac{1}{2}}
\end{eqnarray}

\noindent Using of Eq. (\ref{Double_Total_Derivatives}) for
diagrams with more than one loop of the matter superfields should
be made very carefully. The accurate analysis of the calculations
made in Ref. \cite{Stepanyantz:2014ima} shows that in this case it
is necessary to add terms with insertions of the double total
derivatives into every closed matter loop. For the considered
graph this implies that

\begin{eqnarray}
&&\hspace*{-5mm} \frac{d}{d\ln\Lambda} \frac{\partial}{\partial
\ln g} \Big(\frac{1}{2}\int d^8x\,d^8y\,(\theta^4)_x (\theta^4)_y
\frac{\delta^2\Delta\Gamma}{\delta \mbox{\boldmath$V$}_x \delta
\mbox{\boldmath$V$}_y}\Big) \leftarrow - 8 {\cal V}_4\cdot N_f^2
\sum\limits_{I=0}^n (-1)^{P_I} \frac{d}{d\ln\Lambda} \int
\frac{d^4q}{(2\pi)^4}
\nonumber\\
&&\hspace*{-5mm} \times \Bigg\{ \frac{\partial}{\partial q^\mu}
\frac{\partial}{\partial q_\mu} \Bigg( \frac{1}{(q^2+M_I^2)} \int
\frac{d^4k}{(2\pi)^4} \frac{d^4l}{(2\pi)^4} \frac{e_0^4 g^3}{k^4
R_k^2}\Bigg[ \frac{2}{\left((q+k)^2 + M_I^2\right)}
\sum\limits_{J=0}^n (-1)^{P_J} \frac{1}{l^2 + M_J^2}
\nonumber\\
&&\hspace*{-5mm} + \sum\limits_{J=0}^n (-1)^{P_J}
\frac{1}{\left(l^2 + M_J^2\right) \left((l+k)^2 +
M_J^2\right)}\Bigg]\Bigg) + \frac{1}{(q^2+M_I^2)} \int
\frac{d^4k}{(2\pi)^4} \frac{d^4l}{(2\pi)^4} \frac{e_0^4 g^3}{k^4
R_k^2} \,\frac{\partial}{\partial l^\mu} \frac{\partial}{\partial
l_\mu}
\nonumber\\
&&\hspace*{-5mm} \times \Bigg[ \frac{2}{\left((q+k)^2 +
M_I^2\right)} \sum\limits_{J=0}^n (-1)^{P_J} \frac{1}{l^2 + M_J^2}
+ \sum\limits_{J=0}^n (-1)^{P_J} \frac{1}{\left(l^2 + M_J^2\right)
\left((l+k)^2 + M_J^2\right)}\Bigg] \Bigg\}.\qquad
\end{eqnarray}

\noindent This expression is given by the sum of four terms. The
first two terms (containing the derivatives with respect to
$q^\mu$) are obtained exactly as in the previous section. Two last
terms (with the derivatives with respect to $l^\mu$) are obtained
by inserting the double total derivatives into the closed loop
which corresponds to the momentum $l^\mu$.

Let us now rename the summation indexes and integration variables
in the second and fourth terms by making the replacements $I
\leftrightarrow J$ and $q\leftrightarrow l$. Then we obtain the
expression (\ref{Diagram3}), which was found earlier by explicit
summation of three-loop diagrams with two external lines of the
background gauge superfield.

4. Cutting the matter line in diagram (4) we can obtain a diagram
which is not 1PI. The functions $G_I$ and $J_I$ are evidently
contributed only by the 1PI diagrams, because these functions are
constructed from the effective action. In order to construct all
relevant 1PI diagrams in the considered case it is also necessary
to include 1PI diagrams which are obtained after two cuts of the
matter line. This procedure gives the one-loop diagram presented
in Fig. \ref{Figure_Beta_Diagrams} in the fourth column.
Therefore, calculating the considered contribution we should take
into account both the one-loop and two-loop diagrams presented in
this column. They correspond to

\begin{eqnarray}\label{Matter_Diagrams4}
&& \Delta G^{(4)}_I = -  g^2 \int \frac{d^4k}{(2\pi)^4}\frac{2
e_0^2}{k^2 R_k \left((q+k)^2 + M_I^2\right)}\nonumber\\
&& \qquad\qquad\qquad\qquad - g^4 \int \frac{d^4k}{(2\pi)^4}
\frac{d^4l}{(2\pi)^4} \frac{4 e_0^4 \left((q+k)^2
-M_I^2\right)}{k^2 R_k l^2 R_l \left((q+k)^2 +
M_I^2\right)^2 \left((q+k+l)^2 + M_I^2\right)};\qquad \nonumber\\
&& \Delta J^{(4)}_I=0.
\end{eqnarray}

\noindent Then we calculate the same functions as in Eq.
(\ref{Inverse_Two_Loop}), but keep the terms proportional to $g^4
e_0^4$, which are evidently contributed by both terms in Eq.
(\ref{Matter_Diagrams4}):

\begin{eqnarray}
&&\hspace*{-7mm} \frac{G_I}{4(q^2 G_I^2 + M_I^2 J_I^2)}\leftarrow
\frac{q^2 (q^2- 3 M_I^2)}{(q^2+ M_I^2)^3} \int
\frac{d^4k}{(2\pi)^4} \frac{d^4l}{(2\pi)^4} \frac{g^4 e_0^4}{k^2
R_k l^2 R_l \left((q+k)^2 + M_I^2\right) \left((q+l)^2 +
M_I^2\right)}
\nonumber\\
&&\hspace*{-7mm} + \frac{q^2 - M_I^2}{(q^2 + M_I^2)^2} \int
\frac{d^4k}{(2\pi)^4} \frac{d^4l}{(2\pi)^4} \frac{g^4 e_0^4
\left((q+k)^2 -M_I^2\right)}{k^2 R_k l^2 R_l \left((q+k)^2 +
M_I^2\right)^2 \left((q+k+l)^2 + M_I^2\right)};\quad\nonumber\\
&&\hspace*{-7mm} \frac{ M_I J_I}{q^2 G_I^2 + M_I^2
J_I^2}\leftarrow \frac{q^2(3q^2-M_I^2)M_I}{(q^2 + M_I^2)^3} \int
\frac{d^4k}{(2\pi)^4} \frac{d^4l}{(2\pi)^4} \frac{4 g^4 e_0^4}{k^2
R_k l^2 R_l \left((q+k)^2 + M_I^2\right) \left((q+l)^2 +
M_I^2\right)}
\nonumber\\
&&\hspace*{-7mm} + \frac{q^2 M_I}{(q^2 + M_I^2)^2} \int
\frac{d^4k}{(2\pi)^4} \frac{d^4l}{(2\pi)^4} \frac{8 g^4 e_0^4
\left((q+k)^2 -M_I^2\right)}{k^2 R_k l^2 R_l \left((q+k)^2 +
M_I^2\right)^2 \left((q+k+l)^2 + M_I^2\right)}.\qquad
\end{eqnarray}

\noindent Using these expressions we proceed similarly to the
previous section and find the contribution

\begin{eqnarray}
&&\hspace*{-5mm} \frac{d}{d\ln\Lambda} \frac{\partial}{\partial
\ln g} \Big(\frac{1}{2}\int d^8x\,d^8y\,(\theta^4)_x (\theta^4)_y
\frac{\delta^2\Delta\Gamma}{\delta \mbox{\boldmath$V$}_x \delta
\mbox{\boldmath$V$}_y}\Big) \leftarrow 8 {\cal V}_4 \cdot N_f
\sum\limits_{I=0}^n
(-1)^{P_I} \frac{d}{d\ln\Lambda}\nonumber\\
&&\hspace*{-5mm} \times \int \frac{d^4q}{(2\pi)^4}
\frac{\partial}{\partial q^\mu} \frac{\partial}{\partial q_\mu}
\Bigg( \frac{q^2 - M_I^2}{(q^2+M_I^2)^2} \int
\frac{d^4k}{(2\pi)^4} \frac{d^4l}{(2\pi)^4} \frac{e_0^4 g^4}{k^2
R_k l^2 R_l \left((q+k)^2 + M_I^2\right) \left((q+l)^2 +
M_I^2\right)}
\nonumber\\
&&\hspace*{-5mm} + \frac{1}{(q^2 + M_I^2)} \int
\frac{d^4k}{(2\pi)^4} \frac{d^4l}{(2\pi)^4} \frac{e_0^4 g^4
\left((q+k)^2 -M_I^2\right)}{k^2 R_k l^2 R_l \left((q+k)^2 +
M_I^2\right)^2 \left((q+k+l)^2 + M_I^2\right)}\Bigg).\qquad
\end{eqnarray}

\noindent Making the substitution $q^\mu\to q^\mu-k^\mu$ in the
last integral we obtain Eq. (\ref{Diagram4}).

5. The fifth graph in Fig. \ref{Figure_Beta_Diagrams} is analyzed
completely similarly to the two-loop case. The only corresponding
diagram contributing to the two-point Green functions of the
matter superfields gives

\begin{eqnarray}
&&\hspace*{-5mm} \Delta G^{(5)}_I = - g^4 \int
\frac{d^4k}{(2\pi)^4} \frac{d^4l}{(2\pi)^4} \frac{4 e_0^4
\left((2q+k+l)^2+M_I^2\right)}{k^2 R_k l^2 R_l \left((q+k)^2 +
M_I^2\right)\left((q+l)^2 + M_I^2\right)\left((q+k+l)^2 +
M_I^2\right)}; \nonumber\\
&&\hspace*{-5mm} \Delta J^{(5)}_I= - g^4 \int
\frac{d^4k}{(2\pi)^4} \frac{d^4l}{(2\pi)^4} \frac{4 e_0^4 q^2}{k^2
R_k l^2 R_l \left((q+k)^2 + M_I^2\right)\left((q+l)^2 +
M_I^2\right)\left((q+k+l)^2 + M_I^2\right)}.\nonumber\\
\end{eqnarray}

\noindent Repeating the transformations of the previous section we
obtain Eq. (\ref{Diagram5}).

6. The diagram (6) also contains two matter loops. The
corresponding contributions to the two-point Green functions of
the matter superfields are given by a single diagram presented in
the sixth column of Fig. \ref{Figure_Beta_Diagrams}. This diagram
gives the following contributions to the functions $G_I$ and
$J_I$:

\begin{eqnarray}
&& \Delta G^{(6)}_I = g^4 N_f \int \frac{d^4k}{(2\pi)^4}
\frac{d^4l}{(2\pi)^4} \frac{4 e_0^4}{k^4 R_k^2 \left((q+k)^2 +
M_I^2\right)}\nonumber\\
&&\qquad\qquad\qquad\qquad\qquad\qquad\qquad \times
\sum\limits_{J=0}^n (-1)^{P_J} \Bigg(\frac{k^2 - (k+q)^2 -
q^2}{\left(l^2 + M_J^2\right)\left((k+l)^2 + M_J^2\right)}
- \frac{2}{l^2 + M_J^2}\Bigg);\qquad \nonumber\\
&& \Delta J^{(6)}_I= - g^4 N_f \int \frac{d^4k}{(2\pi)^4}
\frac{d^4l}{(2\pi)^4} \frac{8 e_0^4 q^2}{k^4 R_k^2 \left((q+k)^2 +
M_I^2\right)}\nonumber\\
&&\qquad\qquad\qquad\qquad\qquad\qquad\qquad \times
\sum\limits_{J=0}^n (-1)^{P_J} \frac{1}{\left(l^2 +
M_J^2\right)\left((k+l)^2 + M_J^2\right)}.\qquad
\end{eqnarray}

\noindent Using them we can construct the contribution to the
$\beta$-function defined in terms of the bare coupling constant
according to the prescriptions given above. It is important that
we should also add a term in which double total derivatives are
inserted into the closed loop. The result has the following form:

\begin{eqnarray}
&&\hspace*{-6mm} \frac{d}{d\ln\Lambda} \frac{\partial}{\partial
\ln g} \Big(\frac{1}{2}\int d^8x\,d^8y\,(\theta^4)_x (\theta^4)_y
\frac{\delta^2\Delta\Gamma}{\delta \mbox{\boldmath$V$}_x \delta
\mbox{\boldmath$V$}_y}\Big) \leftarrow -8 {\cal V}_4 \cdot g^4
N_f^2 \frac{d}{d\ln\Lambda} \int \frac{d^4q}{(2\pi)^4}
\frac{d^4k}{(2\pi)^4} \frac{d^4l}{(2\pi)^4} \frac{e_0^4}{k^4
R_k^2} \nonumber\\
&&\hspace*{-6mm} \times \Big(\frac{\partial}{\partial q^\mu}
\frac{\partial}{\partial q_\mu} + \frac{\partial}{\partial l^\mu}
\frac{\partial}{\partial l_\mu}\Big)\sum\limits_{I,J=0}^n
(-1)^{P_I+P_J} \frac{k^2 - (k+q)^2 - q^2 -(l+k)^2 - l^2 - 2M_I^2
-2M_J^2}{(l^2+M_I^2)\left((l+k)^2 +
M_I^2\right)(q^2+M_J^2)\left((q+k)^2 + M_J^2\right)}.\nonumber\\
\end{eqnarray}

\noindent (The term with the derivatives with respect to $l^\mu$
corresponds to insertion of double total derivatives into the
closed loop of the matter superfields.) It is easy to see that
(after an appropriate change of variables) this expression
coincides with (\ref{Diagram6}).

7. The last (two-loop) graph (7) in Fig.
\ref{Figure_Beta_Diagrams} also gives a contribution proportional
to $e_0^2- e^2 \sim e_0^4$. The corresponding functions $G_I$ and
$J_I$ are given by

\begin{equation}
\Delta G^{(7)}_I = - g \int \frac{d^4k}{(2\pi)^4} \frac{2 (e_0^2 -
e^2)}{k^4 R_k};\qquad\quad \Delta J^{(7)}_I= 0.
\end{equation}

\noindent Starting from these expressions by the usual way we
obtain the contribution (\ref{Diagram7}):

\begin{eqnarray}
&& \frac{d}{d\ln\Lambda} \frac{\partial}{\partial \ln g}
\Big(\frac{1}{2}\int d^8x\,d^8y\,(\theta^4)_x (\theta^4)_y
\frac{\delta^2\Delta\Gamma}{\delta \mbox{\boldmath$V$}_x \delta
\mbox{\boldmath$V$}_y}\Big) \nonumber\\
&&\qquad\qquad\quad \leftarrow 4 {\cal V}_4 \cdot g N_f
\frac{d}{d\ln\Lambda} \int \frac{d^4q}{(2\pi)^4}
\frac{d^4k}{(2\pi)^4} \frac{(e_0^2 - e^2)}{k^4 R_k^2}
\frac{\partial}{\partial q^\mu} \frac{\partial}{\partial q_\mu}
\sum\limits_{I=0}^n (-1)^{P_I} \frac{1}{(q^2+M_J^2)}.\qquad\quad
\end{eqnarray}

Thus, we have obtained the correct expressions for all graphs
presented in Fig. \ref{Figure_Beta_Diagrams} without explicit
calculation of the three-loop diagrams with external lines of the
background gauge superfields. This confirms the correctness of Eq.
(\ref{Double_Total_Derivatives}) in the considered approximation.

In order to complete the calculation we derive the NSVZ relation
between the three-loop $\beta$-function and the two-loop anomalous
dimension defined in terms of the bare coupling constant of the
matter superfields. Taking a sum of Eqs. (\ref{Diagram1}) --
(\ref{Diagram7}) and performing the integration over $\ln g$ from
$g=0$ to $g=1$ after simple transformations we obtain

\begin{eqnarray}
&& \frac{\beta(\alpha_0)}{\alpha_0^2} -
\frac{\beta_{\mbox{\scriptsize 1-loop}}(\alpha_0)}{\alpha_0^2} =
4\pi N_f  \sum\limits_{I=0}^n (-1)^{P_I} \frac{d}{d\ln\Lambda}
\int \frac{d^4k}{(2\pi)^4} \frac{d^4q}{(2\pi)^4} \frac{e_0^2}{k^2
R_k} \frac{\partial}{\partial q^\mu} \frac{\partial}{\partial
q_\mu} \frac{1}{(q^2 + M_I^2)}\nonumber\\
&& \times\frac{1}{\left((q+k)^2 + M_I^2\right)} \Bigg\{1 + \int
\frac{d^4l}{(2\pi)^4} \frac{e_0^2}{l^2 R_l} \Bigg(-
\frac{4}{\left((q+l)^2 + M_I^2\right)} +
\frac{2(q^2-M_I^2)}{(q^2+M_I^2)\left((q+l)^2 + M_I^2\right)}
\nonumber\\
&& + \frac{(k+l+2q)^2 + 2 M_I^2}{\left((q+l)^2 +
M_I^2\right)\left((q+k+l)^2 + M_I^2\right)}\Bigg) - N_f
\sum\limits_{J=0}^n (-1)^{P_J} \int \frac{d^4l}{(2\pi)^4}
\frac{2e_0^2}{k^2 R_k}\cdot \frac{1}{(l^2 + M_J^2)}\nonumber\\
&& \times \frac{1}{\left((l+k)^2 + M_J^2\right)} \Bigg\} +
O(e_0^6).\qquad
\end{eqnarray}

\noindent Then we calculate the integrals of double total
derivatives taking into account that they do not vanish only in
the massless case, which corresponds to $I=0$. The result is given
by the following well-defined integrals:

\begin{eqnarray}\label{Beta_Function}
&&\hspace*{-5mm} \frac{\beta(\alpha_0)}{\alpha_0^2} -
\frac{\beta_{\mbox{\scriptsize 1-loop}}(\alpha_0)}{\alpha_0^2} =
\frac{N_f}{\pi}  \frac{d}{d\ln\Lambda} \Bigg\{\int
\frac{d^4k}{(2\pi)^4} \frac{2 e_0^2}{k^4 R_k} + \int
\frac{d^4k}{(2\pi)^4} \frac{d^4l}{(2\pi)^4} \frac{2 e_0^4}{k^2 R_k
l^2 R_l} \Bigg(-\frac{2}{k^2 (k+l)^2}
\nonumber\\
&&\hspace*{-5mm} + \frac{1}{k^2 l^2} \Bigg) - N_f
\sum\limits_{J=0}^n (-1)^{P_J} \int \frac{d^4k}{(2\pi)^4}
\frac{d^4l}{(2\pi)^4} \frac{4 e_0^4}{k^4 R_k^2 (l^2 +
M_J^2)\left((l+k)^2 + M_J^2\right)} \Bigg\} + O(e_0^6).\qquad
\end{eqnarray}

\noindent It is important that the derivative with respect to
$\ln\Lambda$ (which removes infrared divergencies) should be
calculated at a fixed value of the {\it renormalized} coupling
constant $\alpha$.

From the other side, taking the sum of all contributions to the
function $G_{I=0}\equiv G$ and setting $g=1$ we obtain

\begin{eqnarray}
&& G(\alpha_0,\Lambda/q) = 1 - \int \frac{d^4k}{(2\pi)^4} \frac{2
e_0^2}{k^2 R_k (k+q)^2} + \int \frac{d^4k}{(2\pi)^4}
\frac{d^4l}{(2\pi)^4} \frac{4 e_0^4}{k^2 R_k l^2 R_l}
\Bigg(\frac{1}{(q+k)^2 (q+l)^2}\nonumber\\
&& + \frac{1}{(q+k)^2(q+k+l)^2} - \frac{(2q+k+l)^2}{(q+k)^2
(q+l)^2(q+k+l)^2}\Bigg) + N_f \int \frac{d^4k}{(2\pi)^4}
\frac{d^4l}{(2\pi)^4} \frac{4 e_0^4}{k^2 R_k^2 (q+k)^2}\nonumber\\
&& \times \sum\limits_{J=0}^n (-1)^{P_J} \frac{1}{\left(l^2 +
M_J^2\right)\left((k+l)^2 + M_J^2\right)} - q^2 \int
\frac{d^4k}{(2\pi)^4} \frac{2 e_0^4}{k^4 R_k (q+k)^2} \Bigg(
\frac{1}{e_0^2} - \frac{1}{e^2} + 2 N_f \frac{1}{R_k}\nonumber\\
&& \times \sum\limits_{J=0}^n (-1)^{P_J} \int
\frac{d^4l}{(2\pi)^4} \frac{1}{\left(l^2 +
M_J^2\right)\left((k+l)^2 + M_J^2\right)}\Bigg) + O(e_0^6).
\end{eqnarray}

\noindent Except for the last term proportional to $q^2$ all
integrals here do not contain infrared divergences. The last term
vanishes on shell and is finite in the ultraviolet region (in the
limit $\Lambda\to \infty$), because the expression in the brackets
is finite. This term evidently does not contribute to the two-loop
anomalous dimension, because it vanishes after differentiation
with respect to $\ln\Lambda$. That is why it can be omitted (as it
was done, e.g., in \cite{Kataev:2013eta}). The anomalous dimension
defined in terms of the bare coupling constant is given by
\cite{Soloshenko:2003sx}

\begin{eqnarray}\label{Anomalous_Dimension}
&& \gamma(\alpha_0) = - \frac{d\ln
Z(\alpha,\Lambda/\mu)}{d\ln\Lambda} = \frac{d\ln
G(\alpha_0,\Lambda/q)}{d\ln \Lambda}\Big|_{q=0} =
\frac{d}{d\ln\Lambda} \Bigg\{ - \int \frac{d^4k}{(2\pi)^4} \frac{2
e_0^2}{k^4 R_k}
\nonumber\\
&& + \int \frac{d^4k}{(2\pi)^4} \frac{d^4l}{(2\pi)^4} \frac{4
e_0^4}{k^2 R_k l^2 R_l} \Bigg(\frac{1}{k^2(k+l)^2} - \frac{1}{2
k^2 l^2} \Bigg) + N_f \int \frac{d^4k}{(2\pi)^4}
\frac{d^4l}{(2\pi)^4} \frac{4 e_0^4}{k^4 R_k^2}\nonumber\\
&& \times \sum\limits_{J=0}^n (-1)^{P_J} \frac{1}{\left(l^2 +
M_J^2\right)\left((k+l)^2 + M_J^2\right)} + O(e_0^6)\Bigg\}.
\end{eqnarray}

\noindent This expression is well-defined, because due to the
differentiation with respect to $\ln\Lambda$ (which should be made
before the integrations) there are no infrared divergencies.
Comparing Eqs. (\ref{Beta_Function}) and
(\ref{Anomalous_Dimension}) we obtain the relation

\begin{equation}
\frac{\beta(\alpha_0)}{\alpha_0^2} -
\frac{\beta_{\mbox{\scriptsize 1-loop}}(\alpha_0)}{\alpha_0^2} = -
\frac{N_f}{\pi} \gamma(\alpha_0),
\end{equation}

\noindent which gives Eq. (\ref{NSVZ}) with the argument
$\alpha_0$ after substituting the well-known expression for the
one-loop $\beta$-function of the considered theory.

\section{Conclusion}
\hspace{\parindent}

The NSVZ relation in supersymmetric theories is obtained because
the $\beta$-function defined in terms of the bare coupling
constant is given by integrals of (double) total derivatives. In
the Abelian case this statement was proved in
\cite{Stepanyantz:2011jy,Stepanyantz:2014ima} exactly in all
orders. The proof made in Ref. \cite{Stepanyantz:2014ima} is based
on a special identity relating the two-point Green function of the
gauge superfield and the two-point Green functions of the chiral
matter superfields. This relation allows to construct integrals
for the $\beta$-function if only the two-point Green functions for
the matter superfields are known. In this paper the identity
between the Green functions is verified at the three-loop level by
explicit calculations of the Feynman graphs. Starting from the
expressions for the Green functions of the chiral matter
superfields in the two-loop approximation we construct three-loop
integrals for the $\beta$-function defined in terms of the bare
coupling constant. Comparing the result with the sum of the
corresponding supergraphs we see the agreement. This calculation
confirms the results obtained in \cite{Stepanyantz:2014ima}
exactly in all orders, which, in particular, allow to derive the
NSVZ $\beta$-function (for the RG functions defined in terms of
the bare coupling constant) in the case of using the higher
derivative regularization.

\bigskip
\bigskip

\noindent {\Large\bf Acknowledgements.}

\bigskip

\noindent The work of K.V.Stepanyantz was supported by Russian
Foundation for Basic Research grant, project No. 14-01-00695.
K.V.Stepanyantz is very grateful to A.L.Kataev for valuable
discussions.


\end{document}